\documentclass[aps,prl,reprint,superscriptaddress,longbibliography,floatfix]{revtex4-2}
\usepackage{amsmath,amssymb,bm,xcolor,slashed,graphicx}

\usepackage[colorlinks=true,linkcolor=blue,citecolor=blue,urlcolor=blue]{hyperref}
\usepackage{orcidlink}

\newcommand{\MSbar}{\overline{\mathrm{MS}}}
\newcommand{\nf}{N_f}
\newcommand{\eps}{\epsilon}
\newcommand{\dd}{\mathrm{d}}

\begin{document}
\title{Six-Loop Renormalization in Quantum Electrodynamics}

\author{Zeyu Li\,\orcidlink{0009-0008-7005-8343}}
\email{zeyuli@pku.edu.cn}
\affiliation{School of Physics, Peking University, Beijing 100871, China}
\affiliation{Center for High Energy Physics, Peking University, Beijing 100871, China}
\author{Ming-xing Luo\,\orcidlink{0009-0000-5317-5734}}
\email{mingxingluo@csrc.ac.cn}
\affiliation{Beijing Computational Science Research Center, Beijing 100193, China}
\author{Tong-Zhi Yang\,\orcidlink{0000-0001-5003-5517}}
\email{tongzhi.yang@m.scnu.edu.cn}
\affiliation{State Key Laboratory of Nuclear Physics and Technology, Institute of Quantum Matter, South China Normal University, Guangzhou 510006, China}
\affiliation{Guangdong Basic Research Center of Excellence for Structure and Fundamental Interactions of Matter, Guangdong Provincial Key Laboratory of Nuclear Science, Guangzhou 510006, China}
\author{Hua Xing Zhu\,\orcidlink{0000-0002-7129-6748}}
\email{zhuhx@pku.edu.cn}
\affiliation{School of Physics, Peking University, Beijing 100871, China}
\affiliation{Center for High Energy Physics, Peking University, Beijing 100871, China}
\affiliation{Southern Center for Nuclear-Science Theory (SCNT), Institute of Modern Physics, Chinese Academy of Sciences, Huizhou 516000, Guangdong Province, China}

\begin{abstract}
We analytically compute the six-loop corrections to the scale evolution of the electric charge, fermion mass, and fermion field in quantum electrodynamics. The results are obtained in the $\MSbar$ scheme for an arbitrary number of equally charged fermions. They extend renormalization in quantum electrodynamics by one loop order and allow new tests of the behavior of perturbation theory at high orders. The methods developed here also open the way to six-loop computations of renormalization in quantum chromodynamics.
\end{abstract}
\maketitle
\raggedbottom

\textit{Introduction}---The renormalization group (RG) originated in the study of short-distance behavior in quantum electrodynamics (QED)~\cite{Landau:1954nau,Gell-Mann:1954yli}. The beta function and anomalous dimensions describe how the charge, masses, and fields change with the renormalization scale. They determine the ultraviolet counterterms and the logarithmic scale dependence of renormalized Green functions. The analytic four-loop QED beta function was obtained in Ref.~\cite{Gorishnii:1990kd}. The full five-loop result for an arbitrary number of leptons followed from the massless quantum chromodynamics (QCD) vector correlator~\cite{Baikov:2012beta}. Results for a single lepton in several schemes, together with the conversion between the on-shell and $\MSbar$ charges, were obtained in Refs.~\cite{Kataev:2012rf,Baikov:2012rr}.

The six-loop corrections provide a new test of the high-order behavior of perturbative QED, including the size, sign, and flavor dependence of successive terms. Keeping the number $\nf$ of fermion flavors arbitrary separates contributions with different numbers of closed fermion loops and allows comparisons with all-order large-$\nf$ predictions for the beta function~\cite{Palanques-Mestre:1983ogz,Gracey:1996he} and anomalous dimensions~\cite{Gracey:1993mass,Gracey:1994field}.
The calculation also probes the transcendental structure of the RG functions. Individual master integrals can contain a richer set of constants, so their cancellation after renormalization is a nontrivial feature of the calculation. Relations between even-zeta terms at different loop orders~\cite{Baikov:2018nopi,Baikov:2019zmy} constrain this structure. Exact six-loop results make it possible to test these relations and determine the remaining rational and transcendental terms for every power of $\nf$.

Progress in QED renormalization has been closely linked to advances in QCD. Following the discovery of asymptotic freedom at one loop~\cite{Gross:1973id,Politzer:1973fx}, calculations of the QCD beta function were extended to two, three, four and five loops~\cite{Caswell:1974gg,Jones:1974mm,Tarasov:1980au,Larin:1993tp,vanRitbergen:1997va,Czakon:2004bu,Baikov:2016tgj,Herzog:2017beta,Luthe:2017ttg,Baikov:2014qja,Luthe:2017mass,Baikov:2017mass,Luthe:2017ttc,Chetyrkin:2017global}. At six loops, both theories require the reduction and evaluation of five-loop massless propagator integrals after ultraviolet subtraction. High propagator powers and irreducible numerators make this a substantial computational problem. QED has a simpler gauge structure but retains much of the integral complexity of QCD. Developing methods for computing these integrals is therefore an important step toward six-loop QCD renormalization.

In this Letter, we present for the first time the six-loop QED beta function and fermion mass and field anomalous dimensions in the $\MSbar$ scheme for $\nf$ Dirac fermions of equal unit charge. Together with the Ward--Takahashi identity~\cite{Ward:1950xp,Takahashi:1957xn}, they determine the renormalization of the QED fields and parameters in a general linear covariant gauge.

\begin{figure*}[t]
 \centering
 \begin{minipage}[t]{.34\textwidth}
  \centering
  \includegraphics[width=\linewidth]{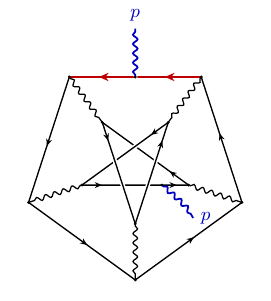}
  \par\smallskip
  {\small (a)}
 \end{minipage}\hspace{.08\textwidth}\begin{minipage}[t]{.34\textwidth}
  \centering
  \includegraphics[width=\linewidth]{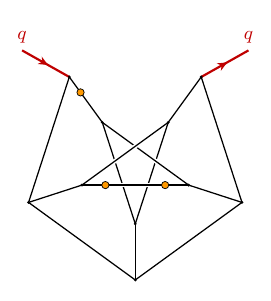}
  \par\smallskip
  {\small (b)}
 \end{minipage}
 \caption{(a) A six-loop photon self-energy diagram contributing to the $\nf^2$ term. Arrows on solid lines indicate fermion flow; wavy lines denote photons. The blue external photons carry momentum $p$. The red propagators adjoining the upper external photon carry momenta $q-p$ and $q$; both receive the auxiliary mass $M$ in the infrared rearrangement. (b) The five-loop scalar propagator family obtained from (a) after expansion in $p$ and infrared rearrangement. The red legs carry momentum $q$ and mark the attachment points of the removed propagators. Each dot denotes an extra propagator power.}
 \label{fig:complexity}
\end{figure*}

\textit{Renormalization}---We use dimensional regularization with $d=4-2\eps$ and the coupling $a=e^2/(16\pi^2)=\alpha/(4\pi)$. The conventional $\MSbar$ scale factor is absorbed into $\mu$. The bare quantities are related to the renormalized ones by $a_0=\mu^{2\eps}Z_a a$, $m_0=Z_m m$, $\psi_0=Z_2^{1/2}\psi$, and $A_0^\mu=Z_3^{1/2}A^\mu$. For the gauge-fixing term $-(\partial\cdot A)^2/(2\xi)$, Landau and Feynman gauges correspond to $\xi=0$ and $1$, respectively. The Ward identities give $Z_a=Z_3^{-1}$, $Z_1=Z_2$, and $\xi_0=Z_3\xi$~\cite{Grozin:2005qedqcd}, where $Z_1$ is the vertex renormalization constant. Only $Z_3$, $Z_2$, and $Z_m$ therefore require independent calculation.

The renormalized parameters depend on the scale $\mu$, while the bare parameters $a_0$, $m_0$, and $\xi_0$ remain unchanged. Differentiating the relations above with respect to $\ln\mu^2$ defines the RG functions:
\begin{align}
 \frac{\dd a}{\dd\ln\mu^2}&=-\eps a+\beta(a), &
 \beta(a)&=\sum_{i\geq0}\beta_i a^{i+2},
 \label{eq:beta}\displaybreak[0]\\
 \gamma_m&=\frac{\dd\ln Z_m}{\dd\ln\mu^2}, &
 \gamma_m(a)&=\sum_{i\geq0}\gamma_m^{(i)}a^{i+1},
 \label{eq:gamma}\\
 \gamma_2&=\frac{\dd\ln Z_2}{\dd\ln\mu^2}, &
 \gamma_2(a,\xi)&=\sum_{i\geq0}\gamma_2^{(i)}(\xi)a^{i+1}.
 \label{eq:fielddefinition}
\end{align}
In this convention, we have $\beta_0=4\nf/3$, $\gamma_m^{(0)}=3$, and $\gamma_2^{(0)}=\xi$. In the $\MSbar$ scheme these functions follow from the single ultraviolet poles of the renormalization constants. The higher poles are fixed by RG consistency relations involving lower-order counterterms~\cite{tHooft:1973mf}.

\textit{The calculation}---We use \texttt{QGRAF}~\cite{Nogueira:1993qgraf} to generate $31\,130$ six-loop photon self-energy diagrams, $31\,130$ six-loop fermion self-energy diagrams, and many lower-order diagrams needed for counterterms. We evaluate the Dirac algebra with \texttt{FORM}~\cite{Ruijl:2017dtg}. The transverse photon self-energy determines $Z_3$. For the fermion self-energy, we retain the vector part $\Sigma_V^0$ and the scalar part $\Sigma_S^0$. To first order in $m_0$, the inverse propagator is $\slashed p(1+\Sigma_V^0)-m_0(1-\Sigma_S^0)$. Finiteness of $Z_2(1+\Sigma_V^0)$ first determines $Z_2$, and finiteness of $Z_2Z_m(1-\Sigma_S^0)$ then determines $Z_m$~\cite{Baikov:2014qja}.

Infrared rearrangement~\cite{Vladimirov:1979zm} allows us to change masses or external momenta while preserving the local ultraviolet counterterm. The infrared poles introduced in this way are removed by the $R^*$ operation~\cite{Chetyrkin:1982nn,Chetyrkin:1984xa,Smirnov:1985yck}. We use its global formulation, which subtracts infrared divergences at the level of entire rearranged Green functions. This approach was developed in earlier correlator calculations~\cite{Chetyrkin:1996scalar,Chetyrkin:1996gluinos} and subsequently extended to five-loop QCD renormalization~\cite{Chetyrkin:2017global,Chetyrkin:2018global}. A complementary diagrammatic formulation for generic tensor numerators is given in Ref.~\cite{Herzog:2017rstar}.

We introduce the same auxiliary mass $M$ into the two propagators adjacent to one external leg. For the photon self-energy, these are the two red fermion propagators in Fig.~\ref{fig:complexity}(a). Expanding in the external momentum yields one-mass tadpoles containing five-loop massless propagator subintegrals. This rearrangement replaces the six-loop propagator problem by the evaluation of five-loop propagators followed by a massive one-loop integration. The global subtraction requires ultraviolet and infrared counterterms, including electron--positron--photon vertex counterterms and a new effective four-photon vertex associated with the infrared rearrangement. The complete subtraction identities and counterterms will be presented in a future publication.

After the global $R^*$ operation, the six-loop photon self-energy alone produces about $2\times10^8$ scalar integrals, which can be organized into $64$ five-loop propagator topologies.  Figure~\ref{fig:complexity}(a) illustrates how one of these families arises: two closed fermion loops connected by five internal photons contribute to the $\nf^2$ part of the beta function. After Taylor expansion in $p$, the red propagators depend only on $q$. Holding $q$ fixed isolates the five-loop family in Fig.~\ref{fig:complexity}(b), with the $q$ integration performed last. At $p=0$, the propagators adjoining the inner blue photon have equal momenta and combine into a squared denominator. The expansion and rearrangement generate further raised powers and irreducible numerators.

This family is among the most demanding families in the calculation, with fourteen denominators and six irreducible scalar products. The dotted integral has total denominator power seventeen and numerator degree seven. We express these integrals in terms of a finite set of master integrals using integration-by-parts (IBP) identities~\cite{Tkachov:1981wb,Chetyrkin:1981qh} and a custom Laporta solver~\cite{Laporta:2000dsw}. We use finite-field sampling and rational reconstruction~\cite{vonManteuffel:2014ixa,Peraro:2016wsq} through \texttt{FiniteFlow}~\cite{Peraro:2019} and organize the reduction by spanning cuts~\cite{Larsen:2016,Wu:2025aeg}. A key ingredient is our custom IBP seeding strategy, designed to efficiently handle the high propagator powers and numerator degrees arising in this calculation. Several public packages which provide automated integral reduction are also available~\cite{Anastasiou:2004air,vonManteuffel:2012np,Lee:2013mka,Smirnov:2019qkx,Smirnov:2023yhb,Smirnov:2025prc,Maierhofer:2017gsa,Klappert:2020nbg,Lange:2025fba,Wu:2023upw,Wu:2025aeg,Guan:2024byi,Ruijl:2017cxj,Chen:2024ampred,Niu:2026secred}. For complementary methods and recent advances on integral reductions, see for example~\cite{Gluza:2010ws,Ita:2015tya,Larsen:2016,Bohm:2018bdy,Smith:2025xes,Guan:2019bcx,Guan:2023avw,Feng:2025leo,Feng:2026imq,Driesse:2024xad,Bern:2024adl,Gehrmann:2026qbl,Song:2025pwy,vonHippel:2025okr,Zeng:2025xbh,Berman:2026jgq}. 

Published expansions of the five-loop massless propagator master integrals reach transcendental weight nine~\cite{Georgoudis:2018olj,Georgoudis:2021}, but they do not provide all $\eps$ orders required here. Poles in the reduction coefficients make higher-order terms in these expansions contribute to the final ultraviolet pole. We therefore evaluate these masters independently using the auxiliary-mass-flow method~\cite{Liu:2017jxz,Liu:2021wks,Liu:2022mfb} with a modified version of \texttt{AMFlow}~\cite{Liu:2023,Huang:2026amflow2}. Our custom IBP and differential-equation solvers provide 400--550 digits of precision. We then reconstruct analytic expressions with PSLQ~\cite{Ferguson:1999} and find agreement with all available results through weight nine.

\textit{Results and checks}---The beta function and mass anomalous dimension are gauge independent. For the fermion field, the Landau--Khalatnikov--Fradkin transformation~\cite{Landau:1955gai,Fradkin:1955jr} fixes the $\MSbar$ gauge dependence~\cite{Grozin:2020}:
\begin{equation}
 \gamma_2(a,\xi)=\gamma_2(a,0)+\xi a,
 \label{eq:fieldgauge}
\end{equation}
where only the one-loop term depends on $\xi$. Therefore, a calculation in Feynman gauge is sufficient to obtain all three RG functions in a general covariant gauge. The corresponding relation $Z_2(a,\xi)=Z_2(a,0)\exp[-a\xi/\eps]$ fixes the gauge-dependent higher poles of the field counterterm. 

The new six-loop coefficients in Eqs.~\eqref{eq:beta}--\eqref{eq:fielddefinition} read 
\begin{widetext}
\begingroup
\setlength{\jot}{3pt}
\setlength{\abovedisplayskip}{5pt}
\setlength{\belowdisplayskip}{5pt}
\begin{align}
 \beta_5={}&\nf\Bigl(-\frac{39\,239}{6}+5888\,\zeta_3-12\,800\,\zeta_5\Bigr)
 \nonumber\displaybreak[1]\\
 &+\nf^2\Bigl(\frac{484\,136}{27}+\frac{149824}{45}\zeta_3-\frac{1024}{3}\zeta_3^2+\frac{1024}{5}\zeta_4+\frac{306880}{9}\zeta_5-\frac{3377024}{45}\zeta_7\Bigr)
 \nonumber\displaybreak[1]\\
 &+\nf^3\Bigl(\frac{87\,760}{27}+\frac{482\,432}{45}\zeta_3-\frac{492032}{45}\zeta_3^2-\frac{27\,968}{15}\zeta_4-\frac{391\,936}{27}\zeta_5+\frac{21\,760}{3}\zeta_6+\frac{71\,680}{9}\zeta_7\Bigr)
 \nonumber\displaybreak[1]\\
 &+\nf^4\Bigl(\frac{209\,104}{243}-\frac{2\,009\,344}{1215}\zeta_3-\frac{4096}{15}\zeta_3^2+\frac{25\,600}{27}\zeta_4+\frac{249\,856}{135}\zeta_5-\frac{10\,240}{9}\zeta_6\Bigr)
 \nonumber\displaybreak[1]\\
 &+\nf^5\Bigl(\frac{16\,064}{3645}-\frac{11\,264}{1215}\zeta_3+\frac{1024}{135}\zeta_4\Bigr),
 \label{eq:b5}\displaybreak[1]\\[2pt]
 \gamma_m^{(5)}={}&-\frac{7\,979\,875}{48}+262\,736\,\zeta_3-52\,992\,\zeta_3^2-482\,280\,\zeta_5+302\,456\,\zeta_7
 \nonumber\displaybreak[1]\\
 &+\nf\Bigl(-\frac{1\,069\,325}{36}-\frac{6\,920\,446}{45}\zeta_3-\frac{409\,856}{3}\zeta_3^2-\frac{2176}{5}\zeta_4
 \nonumber\\
 &\quad-\frac{1\,191\,952}{9}\zeta_5+\frac{20\,800}{3}\zeta_6+\frac{2\,185\,624}{5}\zeta_7-120\,960\,\zeta_9\Bigr)
 \nonumber\displaybreak[1]\\
 &+\nf^2\Bigl(\frac{280\,466}{27}-\frac{13\,147\,504}{135}\zeta_3-\frac{816\,512}{45}\zeta_3^2+\frac{26\,024}{5}\zeta_4
 \nonumber\\
 &\quad-\frac{2\,348\,752}{27}\zeta_5-25\,600\,\zeta_3\zeta_5-\frac{94\,720}{9}\zeta_6+\frac{617\,848}{3}\zeta_7+31\,360\,\zeta_8\Bigr)
 \nonumber\displaybreak[1]\\
 &+\nf^3\Bigl(\frac{359\,090}{243}+\frac{296\,224}{135}\zeta_3-\frac{27\,392}{5}\zeta_3^2-\frac{17\,216}{3}\zeta_4
 \nonumber\\
 &\quad+3584\,\zeta_3\zeta_4+\frac{230\,272}{45}\zeta_5+7040\,\zeta_6-\frac{12\,800}{3}\zeta_7\Bigr)
 \nonumber\displaybreak[1]\\
 &+\nf^4\Bigl(\frac{617\,824}{3645}-\frac{711\,424}{1215}\zeta_3+\frac{4096}{9}\zeta_3^2+\frac{4736}{15}\zeta_4-\frac{5120}{9}\zeta_5+\frac{2560}{9}\zeta_6\Bigr)
 \nonumber\displaybreak[1]\\
 &+\nf^5\Bigl(-\frac{14\,432}{729}-\frac{17\,920}{729}\zeta_3-\frac{2560}{81}\zeta_4+\frac{2048}{27}\zeta_5\Bigr),
 \label{eq:g5}\displaybreak[1]\\[2pt]
 \gamma_2^{(5)}={}&-\frac{2\,359\,327}{144}+\frac{140\,544}{5}\zeta_{5,3}-304\,124\,\zeta_3-245\,840\,\zeta_3^2+145\,408\,\zeta_3^3
 \nonumber\\
 &\quad-\frac{1\,728\,328}{3}\zeta_5-157\,184\,\zeta_3\zeta_5-2\,226\,322\,\zeta_7-\frac{339\,648}{5}\zeta_8+\frac{32\,409\,392}{9}\zeta_9
 \nonumber\displaybreak[1]\\
 &+\nf\Bigl(\frac{23\,463\,211}{540}+\frac{2\,853\,504}{25}\zeta_{5,3}+\frac{11\,966\,528}{45}\zeta_3+\frac{762\,560}{3}\zeta_3^2+10\,240\,\zeta_3^3+29\,600\,\zeta_4+9984\,\zeta_3\zeta_4
 \nonumber\\
 &\quad+\frac{12\,976\,408}{45}\zeta_5-\frac{511\,296}{5}\zeta_3\zeta_5+\frac{245\,200}{3}\zeta_6-694\,596\,\zeta_7-\frac{12\,452\,568}{25}\zeta_8+\frac{1\,574\,336}{9}\zeta_9\Bigr)
 \nonumber\displaybreak[1]\\
 &+\nf^2\Bigl(\frac{65\,188}{15}+\frac{145\,152}{25}\zeta_{5,3}+\frac{4\,159\,792}{135}\zeta_3-\frac{883\,136}{27}\zeta_3^2
 \nonumber\\
 &\quad+2048\,\zeta_3^3-\frac{177\,128}{9}\zeta_4+\frac{47\,872}{3}\zeta_3\zeta_4-\frac{20\,346\,448}{135}\zeta_5-\frac{27\,904}{15}\zeta_3\zeta_5
 \nonumber\\
 &\quad+\frac{1\,158\,400}{27}\zeta_6+\frac{333\,440}{9}\zeta_7-\frac{4\,090\,352}{75}\zeta_8+\frac{1\,228\,864}{9}\zeta_9\Bigr)
 \nonumber\displaybreak[1]\\
 &+\nf^3\Bigl(\frac{260\,846}{243}+\frac{7808}{27}\zeta_3-256\,\zeta_3^2+\frac{2048}{9}\zeta_4-\frac{148\,864}{27}\zeta_5+3200\,\zeta_6\Bigr)
 \nonumber\displaybreak[1]\\
 &+\nf^4\Bigl(-\frac{71\,008}{729}-\frac{512}{27}\zeta_3+\frac{1664}{9}\zeta_4-\frac{1024}{9}\zeta_5\Bigr)
 \nonumber\displaybreak[1]\\
 &+\nf^5\Bigl(\frac{2080}{243}+\frac{2560}{243}\zeta_3-\frac{512}{27}\zeta_4\Bigr).
 \label{eq:h5}
\end{align}
\endgroup
\end{widetext}

Every photon self-energy graph contains a closed fermion loop, so $\beta_5$ has no flavor-independent term. The multiple zeta value $\zeta_{5,3}$ which appears in $
\gamma_2^{(5)}$ is defined by
\begin{equation*}
 \zeta_{5,3}=\sum_{n=1}^\infty \sum_{m=1}^{n-1}\frac{1}{n^5m^3} \simeq 0.0377077 .
\end{equation*}
We also provide ancillary files which contain the complete analytic RG functions and renormalization constants from one through six loops in our conventions.

For $\nf=1$, the numerical expansions read as follows, with coefficients rounded to six significant digits:
\begin{align}
 \beta(a)=&1.33333\,a^2+4a^3-6.88889\,a^4
 \nonumber\\[-2pt]
 &-134.045\,a^5+1506.02\,a^6
 \nonumber\\[-2pt]
 &-32442.5\,a^7+O(a^8),
 \label{eq:beta-nf1}\\
 \gamma_m(a)=&3a-1.83333\,a^2+71.0135\,a^3
 \nonumber\\[-2pt]
 &-1070.30\,a^4+15926.3\,a^5
 \nonumber\\[-2pt]
 &-363199 \,a^6+O(a^7),
 \label{eq:mass-nf1}\\
 \gamma_2(a,\xi)=&\xi a-3.5a^2+6.72222\,a^3
 \nonumber\\[-2pt]
 &+129.606\,a^4-1878.26\,a^5
 \nonumber\\[-2pt]
 &+ 45866.6 \,a^6+O(a^7).
 \label{eq:field-nf1}
\end{align}

We have performed several independent consistency checks. The poles of the six-loop renormalization constants $Z_a$, $Z_m$, and $Z_2$, from $1/\eps^6$ through $1/\eps^2$, agree with the RG predictions obtained from lower-order counterterms. This tests the ultraviolet and infrared subtractions, including the counterterm insertions required by global $R^*$. The new six-loop RG coefficients reside in the remaining single poles.

We further compare the new single-pole coefficients with all-order large-$\nf$ results. The $\nf^5$ term of $\beta_5$ agrees with the result in Ref.~\cite{Palanques-Mestre:1983ogz} and with the abelian limit of the QCD result in Ref.~\cite{Gracey:1996he}. For both $\gamma_m^{(5)}$ and $\gamma_2^{(5)}$, we find agreement for the $\nf^5$ and $\nf^4$ terms with the predictions obtained from critical exponents through $O(1/\nf^2)$~\cite{Gracey:1993mass,Gracey:1994field}. These comparisons test five six-loop flavor coefficients, including their rational and zeta-valued terms. 

We also find complete agreement with the even-zeta relations of Refs.~\cite{Baikov:2018nopi,Baikov:2019zmy} for $\beta_5$, $\gamma_m^{(5)}$, and $\gamma_2^{(5)}$, at every power of $\nf$. These relations determine the even-zeta terms, i.e. $
\zeta_4, \zeta_6, \zeta_4 \zeta_3, \zeta_8, \zeta_4 \zeta_5, \zeta_6 \zeta_3$ from lower-order RG coefficients, testing flavor contributions beyond those fixed by the large-$\nf$ results. The comparison requires expressing $\zeta_{5,3}$ in terms of the weight-eight basis element
\begin{equation}
 \varphi=\frac{3}{5}\zeta_{5,3}+\zeta_3\zeta_5-\frac{29}{20}\zeta_8.
 \label{eq:phi-basis}
\end{equation}
The constrained $\zeta_8$ coefficient is read after rewriting the result in this basis.

We observed that the six-loop beta function has maximal transcendental weight seven, whereas the mass and field anomalous dimensions reach weight nine. Higher-weight constants up to weight 11 occur in the master-integral expansions but cancel after all contributions are combined. Their cancellation provides a further consistency check on the integral reduction and subtraction. 

The coefficients in Eqs.~\eqref{eq:b5}--\eqref{eq:h5} also exhibit a simple arithmetic pattern: after expressing even powers of $\pi$ in terms of even zeta values, every rational coefficient in our zeta basis has the form $p/(2^r3^s5^t)$, where $p\in\mathbb Z$ and $r,s,t$ are nonnegative integers. Thus no prime other than $2$, $3$, or $5$ remains in these denominators.

\textit{Conclusion}---In this Letter, we have analytically determined the six-loop QED beta function and fermion mass and field anomalous dimensions in the $\MSbar$ scheme for arbitrary $\nf$. Agreement with published master-integral expansions, large-$\nf$ predictions, and even-zeta relations, together with the RG consistency of the higher poles, tests the calculation at different stages, from integral evaluation to the final RG coefficients. The calculation required the reduction and evaluation of five-loop propagator integrals with high denominator powers and numerator degrees. Our custom IBP seeding strategy was central to reducing these integrals, including the demanding topology in Fig.~\ref{fig:complexity}. The methods developed here also provide a starting point for six-loop renormalization in QCD, where the same class of propagator integrals occurs with more involved numerator and color structures.

\textit{Acknowledgments}---We thank Zhen-Hua Zhang for his collaboration during the early stages of this project. T.-Z. Yang is supported by the start-up funding from South China Normal University under Grant No. 43102674. M.-x. Luo is supported by the National Natural Science Foundation of China under Grant No. 1249300011. Z. Li and H. X. Zhu are supported by the National Natural Science Foundation of China under Grant No. 12425505. We used large language models (LLMs) to assist in writing this manuscript.

\bibliographystyle{apsrev4-2}
\bibliography{main}
\end{document}